\documentstyle[12pt]{article}

\normalbaselines
\begin{document}
\bibliographystyle{unsrt}
\pagenumbering {arabic}
\vbox{\vspace{38mm}}
\begin{center}
{\LARGE \bf  $D=11$ SUPERMEMBRANE INSTANTONS, \\[2mm]
$W_{\infty}$ STRINGS AND \\ [2mm] THE SUPER TODA MOLECULE  }\\[5mm]

Carlos Castro \\{I.A.E.C 1407 Alegria\\
Austin, Texas 78757  USA }\\[3mm]
( June, 1994. Revised November 1994)\\[5mm]
\end{center}
\begin{abstract}
Exact instanton solutions to  $D=11$ spherical supermembranes moving
in flat target
spacetime backgrounds are construted. Our starting point is
Super Yang-Mills theories, based
on the infinite dimensional $SU(\infty)$ group, dimensionally reduced
to one
time dimension. In this fashion the super-Toda molecule equation is
recovered
preserving only one supersymmetry out of the $N=16$ that one would have

obtained
otherwise. It is conjectured that the expected critical target
spacetime
dimensions  for the
(super) membrane, ($D=11$) $D=27$ is
closely related to that of  the
$noncritical$
(super) $W_{\infty}$ strings.
A BRST analysis of these symmetries should yield information about
the quantum consistency
of the ($D=11$) $D=27$  dimensional (super) membrane. Comments on the
role that
Skyrmions
might play in the two types of Spinning- Membrane actions construted so
far
is presented at the conclusion.
Finally, the importance that integrability on light-lines in
complex superspaces
has in other types of solutions is emphasized.

PACS : 0465.+e;0240.+m
\end{abstract}

\section{Introduction }

Based on the observation that the theory of extended objects beyond the
string
could be seen as a gauge theory of $p$-volume preserving
diffeomorphisms [3] (fo
r
a review) and that supermembranes in $D$ dimensions
(excluding the zero modes), in the lightcone gauge,
are essentially $D-1$ Super Yang-Mills theories dimensionally reduced
to one tem
poral
dimension; we look for solutions to the $N=1~D=10~SDSYM$ of the
$SU(\infty)$
group. i.e; spherical supermembranes moving in flat target space time
background
s.
In $II$ we review the work of [1] in finding special types of solutions
to YM eq
uations in
$R^{4n}$ which is crucial in the construction of the membrane instanton
solution
presented in $III$. This solution preserves only $one$ supersymmetry
and enables
 us
to
arrive at the $SU(\infty)~D=1$ Toda molecule which bears strong
resemblance with
a higher-conformal-spin-extended version of Liouville-like field theory

(dimensionally reduced to one dimension ).
The geometrical origin of $W_{\infty}$ symmetries in $noncritical$
string theory is outlined in $IV$
and a plausible connection between the spectra of
$noncritical~W_{\infty}$ strin
gs
and membranes is raised. A BRST analysis reveals that the target
spacetime
of noncritical $W_{\infty}$ strings is closely related to the expected
critical dimensions for the membrane. $D=27$ for the bosonic membrane
and
$D=11$ for the supermembrane. Quantization could be implemented through
the Quan
tum
Group programm based on the $U_q [SU(\infty)]$ Quantum Group. Finally,
the relev
ance
that
integrability on light-lines in complex superspaces might have in
the construction of more general solutions
is emphasized. An appendix is included where the $N=\infty$ limit of
the Toda
molecule is taken and whose solutions can be obtained in terms of
solutions to
the effective $3D~sl(\infty)$ continuous Toda equation. The latter is
equivalent
to a rotational Killing symmetry reduction of Plebanski's first
heavenly equatio
n
in $D=4$.

\section{The Yang-Mills Equations in $R^{4n}$}

We are going to follow closely [1].
Solutions of the equations for classical YM theory in Euclidean space
$R^{4n}$ for the YM potentials $A_a,~a=1,2,....4n$, and
field-strengths,
$F_{ab}=\partial_a A_b - \partial_b A_a + [A_a,A_b]$,
with values in the semisimple Lie algebra $\cal G$
$$ \partial_a F_{ab} +[A_a,F_{ab}]=0. \eqno (1)$$
can be obtained from the ansatz [1]:
$$A_a= -J^\alpha_{ac}T_\alpha (\phi) \partial_c \phi.  \eqno (2) $$
with the real antisymmetric tensors $J^\alpha_{ab}$ satisfying :
$$J^\alpha_{ac}J^\beta_{bc}=\delta^{\alpha \beta}\delta_{ab} +
\epsilon^{\alpha \beta \gamma}J^\gamma_{ab}. \eqno (3)$$
where $ \epsilon_{\alpha \beta \gamma}$ are the $SU(2)$ structure
constants and
$\alpha ,\beta ,\gamma =1,2,3.$ With the introduction of these tensors
one
introduces a quaternionic strcuture in $R^{4n}=H^n$.
Upon substitution of (2) and using the relations (3) one obtains the
system
of differential equations  :
$$(\partial T_\alpha /\partial \phi) +
{1\over 2}\epsilon_{\alpha \beta \gamma} [T_\beta ,T_\gamma].=0.\eqno
(4)$$
These are the Rouhani-Ward equations [1] and
$$2J^\alpha_{ac}\partial_c \partial_b \phi -
2J^\alpha_{bc}\partial_c \partial_a \phi +
2\epsilon^\alpha_{\beta \gamma} J^\beta_{ac} J^\gamma_{be} \partial_c
\partial_e \phi +J^\alpha_{ab} \Box \phi =0 \eqno (5)$$.

One must emphasize that $T_\alpha =T^A_\alpha L_A$ are $\cal G$-valued
objects whose Lie algebra bracket is $not$ necessarily that of $SU(2)$.
One can also have more general $f_{\alpha \beta \gamma}$ in (4)
belonging
to ${\cal H}$. In the membrane's case $T_\alpha $ become a set of three
$c$-numb
er functions
depending on the two extra internal spatial variables of the membrane
and of the
$\phi$ function which depends on the time variable $x_0$ which is in
$correspondence$ with the  membrane's clock $\tau =X^+=2^{-1/2}(
X_0+X_{10})$. The gauge field of area-preserving diffeomorphisms of the
two
dimensional spherical membrane $\Sigma$ with two spatial coordinates
$\sigma_1,\sigma_2$ is replaced as:
$\omega^{12} \Rightarrow A_0$ which is the extra YM field needed in
order to hav
e
a $D=10~YM$ theory of the $sdiff~\Sigma$ group dimensionally reduced to
one
time dimension.

Having written down the eqs-(4,5) in $R^{4n}$ allows us to relate them
to a
relative set of equations in $R^4$ as follows  :
Intoduce the tensors $J^{\alpha} _{(\mu i)(\nu j)} =
\delta_{ij}\eta^{\alpha}_{\mu \nu}$ with the double-index notation :
$\mu ,\nu ,..=1,2,3,4.~i,j.....=1,2,.....n$ so that
$a,b,...=1,2,.......4n$.
$\eta^{\alpha}_{\mu \nu} $ are the 't Hooft tensors : three real
antisymmetric
$4\times 4$ matrices with components :$\eta^{\alpha}_{\beta \gamma} =
\epsilon ^{\alpha}_{\beta \gamma }$ and
$\eta^{\alpha}_{\mu 4} =-\eta^{\alpha}_{4\mu} =\delta^{\alpha}_\mu .$
Substituting these tensors in (5) one gets the equations  :
$$\partial _{\mu i} \partial_{\nu j} \phi =
\partial _{\mu j} \partial_{\nu i} \phi .\eqno (6a) $$
$$\partial_{\lambda i}\partial _{\lambda j}\phi =0. \eqno (6b)$$

The equation $\Box \phi =\partial_{\lambda j}\partial_{\lambda j} \phi
=0$
follows from (6b) where one sums over the $j =1,2......n$ indices only.
Having solutions to eqs (4,5,6) will yield solutions of the original YM
eqs (1)
after using (2). Now that we have presented a brief review of [1] we
turn
to the membrane's  equations.

\section {The Membrane Instantons and the Toda Molecule}.

A close up study reveals that the $D=11$ supermembrane admits exact
octonionic
and quaternionic instanton solutions which are tightly connected to the
super
Toda molecule equation [4] associated with the minimal embedding of the
$SU(2)$
group into the $N\rightarrow \infty$ limit of $SU(N)$. A special class
of solutions can be found as follows if one
chooses the following ansatz:
In $D=10$ take a $10\times 10~J^\alpha_{AB}$ matrix in block form :
$J^{\alpha}_{MN}\otimes {\cal I}^{\alpha} $ where $J^\alpha_{MN}$ is a
matrix
of the type in (3) with $M,N..=0,1,....7$.
${\cal I}^\alpha $ is a set of three arbitrary  $2\otimes 2$
matrices for the values of $\alpha ,\beta ,..=1,2,3;$  and whose matrix
indices
range over $8,9$ only.
Due to the fact that the membrane is, in effect, a dimensional
reduction to one
dimension of a $D=10$ YM theory, the two YM potentials $A_8, A_9$ are
going to
be zero after a direct application of the ansatz in (2) and due to
the block
diagonal form of $J$. i.e; $A_8 \sim J_{08}\partial_{x_0} \phi =0$.
The same applies with $A_9\Rightarrow 0$. One is not constraining
$a~priori$ the
se
two YM potentials $A_8 ,A_9 $ to zero; these are zero a posteriori as a
result of the dimensional
reduction to one dimension (essentially).
The
$8\otimes 8~J^{\alpha}_{MN}$ matrices are of the form :

$\delta_{ij}\eta^{\alpha}_{\mu \nu}$ as explained in the previous
section. One
could opt to extend the range of indices for $\alpha ,\beta ...$ beyond
$1,2,3$
to four indices, instead, and choose for the four $J$ matrices
a selected subset of the seven antisymmetric $8\otimes 8$ matrices
involved in the Clifford algebra
$Cl(O,6)$ with generators $\gamma^1 ,\gamma^2,....,\gamma^6 ;~\gamma^7
=
\gamma^1 . \gamma^2 ......\gamma^6 $. See [1] for further details where
they
discuss that the range of indices compatible with the ansatz in (2)
can only be those whose $\alpha,...$ are $4,5,6,7$. The sort of
equations one
gets is an octonionic analog of the Rouhani-Ward-Nahm equations :
$$f_{\alpha \beta \gamma} (\partial T_\gamma / \partial \phi )
 +[T^a_\alpha G_a ,T^b_\alpha G_b ] =0.                     \eqno
(7a)$$
where the function $\phi (X_M)$ obeys an equation like (5).  There is
also another alternative way to solve the effective YM in $d=8$;
although the number
of equations increases dramatically, see Popov [6]. These authors
projected an arbitrary
antisymmetric tensor $F_{AB}$ into the orthogonal $21$ and $7$
dimensional subspaces
of the $28$-dim vector space of antisymmetric tensors in $d=8$ given
by the self dual and antiselfdual parts of the tensor $F_{AB} =F^+
+F^-$.

The projectors involved completely antisymmetric tensors built entirely

from the octonionic
structure constants (Cayley numbers) and the standard $\epsilon ,
\delta$ tensor
s.
The value of $\phi =1+X_M X_M $ is the one which allows to recast the
SD equations, for the adequate pertaining ansatz :

$$A^a_M G_a =(-1/6)G_{MNCD}X_N W_{CD} (\phi ) \eqno (7b)$$
to acquire the form :
$$[W_{AB} ,W_{CD} ] =S_{ABCDMN} (\partial W_{MN} /\partial \phi). \eqno
(7c)$$
with $S_{ABCDMN}$ the $SO(8)$ structure constants.
Similar type of octonionic SDYM equations have been discussed in [5].
We are not going to solve these equations and their octonionic
superspace
generalizations because there is a simpler way to reach to the
super-Toda molecule
without having extended supersymmetries in $D=1$ and without to have to
truncate
the dimensionally-reduced ten dimensional YM theory .
We borrow our results in [4] by first reducing the $d=8$ YM to $d=4$ as
shown
in section $II$ and then concentrating on the $d=4$ YM theory with YM
potentials
$A_0 ,A_1 ,A_2 ,A_3$ depending solely on one coordinate which is the
$x_o$ coordinate.

Because of YM gauge invariance one has
enough freedom to choose the gauge $A_0 =0$.
This is due to the fact that  $A_0$ is the relative of the gauge field
of $sdiff
{}~\Sigma$.
[3]. The YM equations in $d=4$ for this choice of gauge and for the
case that
the fields depend solely on one coordinate become :
$$ A_{i,oo} -[A_j ,[A_i ,A_j ]] =0.~~~[A_{i,o} ,A_i] =0. \eqno (7d)$$
where $i,j,k ..=1,2,3; $ only . Eqs.(7d) are in fact equivalent
to the SDYM in $D=4$ :

$$\epsilon _{\mu \nu \rho \zeta } F_{\rho \zeta} =2F_{\mu \nu}. \eqno
(8a)$$

Clearly, when the four indices for the $ \epsilon$ tensor have only
values ranging
between $1,2,3$ one gets zero and when they range over $0,1,2,3$ one
gets
$\epsilon_{ijk}$ so the end result is
$$\epsilon_{ijk}A_{k,o} +[A^a_i G_a ,A^b_j G_b] =0. \eqno ( 8b )$$
For a more general discussion of the equivalence between (7) and (8)
see [1].
In the membrane'case one replaces Lie-algebra brackets with Poisson
brackets.
The semicolons indicate that the derivatives are taken with respect to
$x_o$.

{}From [1] : $x_0 =p_1 x_{01} +p_2 x_{02}$ so that
$(\partial /\partial x_{01}) =p_1 (\partial /\partial x_0)....$ and the
potentia
ls
are obtained for $\phi (x_0) =x_0$ :
$${\cal A}_6 =p_2 \eta^\alpha_{20} A_\alpha (x_0 ; \sigma^1, \sigma^2 )
=
 p_2A_2 (x_0 ,\sigma^1 ,\sigma^2).~{\cal A}_2 =p_1A_2. $$

$${\cal A}_1 =p_1A_1.~{\cal A}_5 =p_2A_1.~
{\cal A}_3 =p_1A_3.~{\cal A}_7 =p_2A_3.$$
$${\cal A}_0 =p_1\eta^\alpha_{00}A_\alpha=p_1A_0= {\cal A}_4 =0 \eqno
(8c) $$
The last equation is consistent with the gauge choice $A_0 =0$.

Equations (7a) in the $R^{4n}$ case, and for the special choice of the
$J$ matri
ces
discussed in section $II$,  have a Lax-type representation in terms of
a spectral parameter $\lambda$. Applying the inverse scattering
transform
method
allows to construct solutions of Lax's equation in terms of theta
functions
for any semisimple

Lie algebra $\cal G$ and can be reduced to Toda
lattice equations [6]. On the other hand, eqs- (7a,7b) in $d=8$ for the

special case
that
the Lie algebra ${\cal G} \equiv {\cal H}$
may be obtained from classical Yang-Baxter equations. For
simple Lie algebras these admit three classes of solutions
: elliptic, trigonometric and rational
solutions [6].
However, this is $not$ necessary for the $SU(\infty)$
group when one uses Poisson brackets instead.
Initially, the authors [7] expressed solutions of the
self dual membrane in $D=4+1$ in terms of solutions of the Toda
molecule for
$SU(2)$
after an appropiate $truncation$ of the YM potentials in the expansion
of spheri
cal
harmonics. In [4] we  found solutions of the self dual supermembrane
in $4+1$ $without$ a truncation. The ansatz which allows to recast
 the $SU(\infty)$ Nahm's equations as a super-Toda molecule is :
$$\{A_y ,A_{\bar y}\} =-i ~\sum^{\infty}_{l=1}~exp
(K_{ll'}\Phi_{l'})Y_{l0}
(\sigma_1 ,\sigma_2). \eqno (9a)$$
and :

$$A_3 =-\sum^{\infty}_1~(\partial \Phi_l /\partial \tau) Y_{l0}. \eqno
(9b)$$
where $A_y \sim A_1 +iA_2.~A_{\bar y} \sim A_1 -iA_2$.
and $A_y =\sum~A_{yl} (x^0_L ,\theta^+) Y_{l, +1}(\sigma_1 ,\sigma_2
)$.
$A_{\bar y}$ is expanded in terms of $Y_{l,-1}$ and $A_3$ in terms of
$Y_{l,0}$.

Taking $\partial A_3/\partial \tau =i \{A_y ,A_{\bar y} \}$ gives :

$$-{\partial^2 \Phi_l \over {\partial  ( x^0_L )^2 }} =exp (K_{ll'}
\Phi_{l'}).
\eqno (10)$$
where $\tau =x^0_L$.
This is the $SU(N)$ super-Toda molecule equation in Minkowski form for
the $\Phi_1,
\Phi_2 ,\Phi_3 ,......$ superfields where the $SU(2)$ has been
$minimally$ embedded
into $SU(N)$. This explains the presence of the spherical harmonics in
(9).
$x^0_L$ is a left-handed coordinate and
${\theta}^+ $ is the light-cone chiral Grassmanian variable [8].
The anti self dual YM equations require the use of anti-chiral
light-cone
superfields $ {\bar \Phi} (x^0_L, {\bar \theta}^+)$.

$K_{ll'}$ is the $SU(N)$ Cartan matrix.
$\Phi_1 ,\Phi_2 ,.....$ are light-cone chiral superfields in Lorentzian
superspace encoding the true physical propagating degress of freedom
of YM in $3+1$
dimensions [8]. Gilson et al by integrating out the inappropiate
light-cone projections
of the superspace Grassmanian variables obtained a SDSYM formulation
with only
physical propagating modes. Auxiliary fields and nonpropagating modes
were
completely eliminated. The end results are not manifestly Lorentz
covariant
and are supersymmetric under the light-cone supersymmetry
$Q_{+\alpha}$.
This light-cone chiral superfield formulation is compatible with the
initial light-cone supermembrane action insofar as the light-like
directions
of the $D=11$ supermembrane is concerned. There is a $correspondence$
(not an strict identification
! ) between the membrane's coordinates and the YM potentials.
Also, the membrane's clock
is $X_+ =X_0+X_{10}$ which $corresponds$ to the $x_o$ coordinate of the
$D=10$
YM theory.
If one $identifies$ $X_+$ with $x_0$ which, in turn, is seen as $X_0$,
one would
have to constrain the membrane's $X_0 ,X_{10}$ coordinates !
The value of the left handed coordinate is therefore :
$x^0_L \equiv x^0 +i\theta \sigma^0 {\bar \theta}$.
The fact that only half of the supersymmetries are linearly realized
in the $D=11$ light-cone supermembrane [3] is also consistent with the
fact
that the light-cone chiral superfield formulation has only
manifest supersymmetry under $Q_{+\alpha}$.

Furthermore, a naive dimensional reduction of the $N=1~D=3+1~SDSYM$
theory
to $D=1$ would have generated $N=4$ supersymmetries.
And, as it is known, a naive dimensional reduction of $N=1~D=10~SYM$
yields $N=16~D=1~SYM$.
Therefore, the light-cone chiral-superspace formulation generates only
$one$
supersymmetry out of the $N=16$ . Octonionic superstring soliton
solutions
to the low-energy heterotic-field-theory equations of motion in $D=10$
preservin
g
only $one$ supersymmetry out of the sixteen were found in [9]. The
soliton
described a ten dimensional cosmic (not fundamental) string in
Minkowski space acting as a source for massless
spacetime fields transverse to the two dimensional world of the string.
Since
the transverse space of the $d=2+1$ supermembrane in $D=10+1$ is the
same as that of
the $d=1+1$ superstring in $D=9+1$ it is not surprisising that
similarities
occur. Also the role of octonions in the construction of the
exceptional group
$E_8$ and the triality properties of the transverse compact group
$SO(8)$ hint to
to important connections between the two. Dualities between strings and
five-branes
seem to point out that these two are dual description of the same
physics [10].
It is warranted to explore all these connections deeper and see what
new physical results are encountered.
It was emphasized [4] that although the Lorenztian superspace solutions
of the SDSYM equations are $complex$ valued ( which would also
force the membrane's coordinates to be as well)- after the dimensional
reduction has taken place
from $4 \rightarrow 1$ a $reality$ condition on the complex superfields
can be met [8]. This is consistent with the fact that $N=1~D=4$
Euclidean supersymmetries
can only be analytically continued to Minkowski space iff the number of
Minkowski
supersymmetries is $N=2$. There are no Majorana spinors in four
Euclidean dimensions but there are in ten and eight Euclidean [11],
which was
our starting point to construct instanton solutions. At the end of the
road in
Euclidean $D=1$ a Majorana spinor condition can be imposed.
Whereas in Minkowski
$D=1$ space a pseudo-Majorana spinor exists.
In the appendix we show that the $N\rightarrow \infty$ limit of (10)
(the bosonic piece) is nothing but a dimensional reduction of the $3D$
continuous Toda equation [37] given by Leznov and Saveliev :
$$\partial_z \partial_{\bar z} u =-\partial^2_t (e^u).\eqno (11)$$
where $z,{\bar z}$ are complex coordinates and $t$ is a continous
parameter. A dimensional reduction from $3D$ to $2D$ yields :
$$\partial^2_r u =-\partial^2_t (e^u).~~r\equiv z+{\bar z}. \eqno
(12)$$

Solutions to (12) have been given by Saveliev [37] in terms of an exact
series expansion of the quantity $\mu ={d(t)e^{r\phi(t)}\over
{(\phi (t))^2}}$ where $d(t)$ and $\phi (t)$ are arbitrary functions of
$t$
:
$$e^x =e^{x_o}[1-\mu +1/2 \mu^2 +......]. \eqno (13a)$$
where $x$ is a solution of
$$\partial^2_r x =e^{\partial^2_t x}. \eqno (13b)$$
$$\partial^2_t x_o =r\phi +lnd. \eqno (13c)$$
where one has set the coupling constant to one and $x_o$ is an
asymptotic
solution at ($r=\infty$). For more details see [37]. An exact
quantization of
the Toda lattice has been given by Leznov and Saveliev in chapter seven
of
their book [38]. In principle a quantization of the continous Toda
theory could
be achieved by taking the continuum limit in all of their equations.
After quantization the solutions (13) will receive $h$ corrections.
There
is no assurance that these limits would behave smoothly and probably
renormalization would be required. A Quantum Group approach certainly
is
very useful in order to evalute the operator products of  Liouville-
type fields exponentials
however as far as we know the Quantum Group associated with the
$SL(\infty)$ algebra has not been constructed. Saveliev gave an
explicit
realization of the $W_{\infty}$ algebra in terms of the $3D$ continous
Toda
field since the continous Toda exhibits a $W_{\infty}$ symmetry. Quasi
finite Unitary Highest Weight representations of the $W_{\infty}$
algebra have been
given in [39. Thus the quantum spectrum of the continous Toda theory
can
in  principle be computed. We won't go into these details but instead
we
suggest the connection to noncritical $W_{\infty}$ strings in the next
section.

\section{ Anomalies, Spectrum and Light-like Integrability}

Using the tecniques and results [6] of constructing solutions of the
Toda
lattice and Yang-Baxter equations would facilitate
the construction of solutions to (10) and a subsequent
quantization process will provide for us the spectrum of the
supermembrane.
The authors in [12] gave a rigorous proof that the Hamiltonians of
Supersymmetric Gauge Quantum Mechanical Models (SGQMM) for finite
matrices  have a continuous spectrum starting at zero
and, in particular, with no mass gap.
For this reason they concluded that the supermembrane was
unstable against deformations
into long stringlike configurations of zero area ( zero energy). In the
bosonic
case quantum effects saved the day and prevented these instabilities to
occur
[13]. Nevertheless, [12] emphasized that matters could change in the
$N=\infty$
limit. Mainly, this limit is $not$ unique : it is basis dependent.
Therefore
discontinuities do in fact occur. For this reason one cannot be
absolutely certain
that results for the finite $N$ truncation of the supermembrane apply
as
well for
the full fledged supermembrane theory. It is for this reason that
solutions to the super-Toda molecule equation and their subsequent
quantization
will in principle provide for us the long-sought spectrum of the
supermembrane.
Supersymmetry $is$ present in (10) without invoking to Witten
index evaluations nor
worrying if it makes sense to evaluate it in
the continuous
spectrum case and whether is infinite or not.

The underlying origins of $W_{\infty}$
symmetries of the continuous Toda equation is closely related to the
$W_{\infty}$ symmetry  in $W_{\infty}$ strings as we shall see below.
These stem
 from $D=4~SU(\infty)$ SDYM as follows :

In [15] a geometrical meaning to $W_{\infty}$
gravity as a gauge theory of volume-preserving diffs in the space of
dimensionally-reduced $SU^*(\infty)$ instantons was presented.  The
origins of a universal linear and nonlinear $W_{\infty}$
algebras was provided in terms of the geometry of the $SU^*
(\infty)~SDYM$ in $D=4$ ( An infinite dimensional moduli space).
A dimensional reduction gives SD gravity in $D=4$ (Plebanski's second
heavenly
equation). A Darboux change of variables converts it into the first
heavenly equation and a further Killing
symmetry reduction furnishes the $sl(\infty)$ continuous Toda equation.

The latter is
an effective $D=3$ equation whose symmetry algebra contains the linear
classical
$w_{\infty}$ algebra [16,37].

A Moyal deformation of the Poisson bracket yields the
centerless $W_{\infty}$ algebra [17].
Central extensions are obtained by means of the
cocycle formula in [17]. Central terms of this sort were also
considered in term
s of the
symmetries of the $tau$ function associated with the $sdiff~\Sigma$
group
by [18].
A $nonlinear$ bracket based on $nonlinear$ gauge theories of the Kyoto
school
was developed in [15] which made contact with the nonlinear
$W_{\infty}$ algebras of [19]. Also, the mapping of solutions
( after a dimensional reduction) from Plebanski equations
to those of the KP equation was made in [15] and the role of many
different types of
$W_{\infty}$ algebras as subalgebras of the KP equation was  found.
Since the KP hierarchy contains many
of the known integrable hierarchies of nonlinear differential equations
in
lower dimensions it is not surprising that the Toda molecule
(a generalization of the
dimensionally reduced Liouville-like theory) appears within this
context
because we had started from a supermembrane which is a gauge theory of
the
$sdiff~\Sigma \sim SU(\infty)$ group.
In view of this fact it is very plausible that the spectrum of the
super-Toda
molecule
is related  to the $infinite$ tower of massless
higher-superconformal-spin analo
gs of the
super-Liouville modes ( $after$ a dimensional reduction) present in
$noncritical
$linear and nonlinear $ W_{\infty}$ strings. Therefore,
the membrane's spectrum (from the point of view of
the target spacetime)
might be related to that of the $noncritical~  W_{\infty}$
strings moving in flat target spacetimes. This is worth looking into.
The presence of the Liouville-like modes signals strings in
non-critical
dimensions. The regularized central charge for critical linear
$W_{\infty}$
strings was $c=-2$ [28]. Clearly one cannot have a membrane moving in
negative spacetime dimensions. Since the Liouville modes are present
then
$c_m +c_L +c_g =0 \Rightarrow c_m \neq -2$.
Furthermore, the vast literature on Quantum Groups should provide
insights as to
how quantize the membrane : the Quantum Liouville theory has been
extensively
studied from many points of view ; mainly in the continuum limit,
matrix models
....

[29].
The group to be studied is :$U_q [SU^* (\infty)] \sim U_q [SL[(\infty ,
H)]$.

This is what we are going to do now : Establish the correspondence
between
the target spacetime of noncritical $W_{\infty}$ strings and the
membrane's
embedding coordintes and find that $D=27;11$ -the expected number of
spacetime
dimensions for the  critical bosonic membrane and supermembrane -can be
accomodated within the context of a $W_{\infty}$ conformal field
theory..

$W$ symmetries of Toda models have been known for sometime. The Toda
theory can
be obtained
from a constrained $WZNW$ model [31]. Also, the induced action of $W_n$
gravity
in
the conformal gauge takes the form of a Toda action for the scalar
fields and the $W$
currents take the familiar free field form [32]. Each of the Toda
actions posseses a
$W_n$ symmetry. Furthermore, $W_\infty$ gravity in the light-cone gauge
posesses an underlying $sl(\infty)$ Kac-Moody symmetry [33] which is to
be contrasted
with the fact that $sl (\infty)$ Toda theory has $w_\infty$ for
underlying
symmetry. A Moyal deformation and the use of the nonlinear bracket [15]
yields
the linear and nonlinear $W_\infty$ algebras.

Our aim is to show that the spectrum of (10) could be classified in
terms
of representations of $W_\infty$ algebras [38]
and the instanton sector of the membrane (excluding the zero modes)
hereby discussed is closely
related to
the spectrum noncritical $W_\infty$ strings.

A BRST quantization of the continuous Toda action will encounter
anomalies
in the quantum $W_{\infty}$ algebra as a result of normal ordering
ambiguities, per example, as it occurs in the string. This would
destroy
unitarity in the spectrum present in the physical Hilbert space of
states and
hence full Lorentz
invariance of the target spacetime will not hold . Pope et al [34] have
shown that an
anomaly-free quantum theory can be constructed if the "matter"
realization of
the $W_{\infty}$ algebra has for central charge a regularized value of
$c_m=-c_{gh}=-2$ in order to have a nilpotent BRST charge operator. If
this
is so the quantum theory will be devoid of universal gauge anomalies.
The $W_{\infty}$ algebra involved here is the $W$ algebra associated
with
the Lie algebra of $SU(\infty)$ and this is the $A_{\infty}$.

$W_{\infty}$
string is a generalization of ordinary string theory in the sense that
instead of gauging the two-dimensional Virasoro algebra one gauges the
higher conformal-spin algebra generalization :the $W_{\infty}$ algebra.
The
spectrum can be computed exactly [35] and is equivalent to a set of
spectra
of Virasoro strings with unusual central charges and intercepts. In
particular, the critical $W_N$ string
( linked to the $A_{N-1}$ algebra ) has for central charge the value :
$$c=26 -(1-{6\over {N(N+1}}).\eqno (14)$$
for which unitarity is achieved if the conformal-spin two-sector
intercept
is
$$\omega_2 = 1-{k^2-1\over {4N(N+1)}}.\eqno (15)$$
$k$ is an integer ranging between $1$ and $N-1$.

The role of the unitary Virasoro minimal models in the expression for
the
central charge given by (14)
was explained  and clarified by [29] when they constructed in general
the BRST operator
corresponding to a $W_N$ algebra as a nested sum of nilpotent and
mutually
anticommuting BRST operators. This required  a new basis in the Hilbert
space. In this fashion the exact cohomology can be calculated by an
iterative procedure.

In the noncritical $W_N$ string case [29] the matter and
Liouville sector of the $W_N$ algebra can be realized in terms of $N-1$
scalars, $\phi_k;\sigma_k$, respectively. These realizations in general
have background charges which are fixed by the Miura transformations
[29,36]. The noncritical $W_N$ string is characterized by the central
charges of the matter and Liouville sectors $c_m,c_L$ respectively. To
achieve a nilpotent BRST operator these central charges must satisfy :
$$c_m+c_L =-c_{gh}=2\sum^N_{s=2}~(6s^2-6s+1)=2(N-1)(2N^2+2N+1).\eqno
(16)$$
Where one has summed over all the conformal spins $s=2,3.....N$. In the
$N\rightarrow \infty$ limit one has to perform a zeta function

regularization [34] scheme which yields the value of $-2$ for the right
hand side of (16).
The authors in [29] have shown that the nested basis can be chosen
either
for the Liouville sector or the matter sector but $not$ both. [29]
chose
the nested basis for the Liouville sector  and
found that :
$$c_L =(N-1)[1-2x^2N(N+1)].\eqno (17)$$
where $x$ is an arbitary  parameter which makes it possible to avoid
the
relation with the minimal models. By choosing $x$ appropriately one can
of
course get the $q$ th unitary minimal models by having
$$ x^2_o=-2-{1\over 2q(q+1)}.\eqno (18)$$
where $q$ is an integer. In this case from (16,17) one gets for the
central
charge of the matter sector :
$$c_m =(N-1)(1-{N(N+1)\over {q(q+1)}}). \eqno (19).$$
which corresponds to the $q$ th minimal model of the $W_N$ string.
In the present case one has the freedom of selecting the minimal model
since the value of $q$ is arbitrary. If one chooses $q=N$ then $c_m=0$
and
the theory effectively reduces to that of the "critical" $W_N$ string.
We are now approaching the main point of this work. In order to find a
spacetime interpretation, the coordinates $X^\mu$ must be related to
the
single scalar field of the Liouville sector $\sigma_1$ (since one
decided to choose
the nested basis in the Liouville sector) appearing in the Miura basis
as well as in the
nested basis. It happens that  the stress energy momentum tensor is not
modified when one
performs a field redefinition from the Miura to the nested basis [29].
The
other higher spin currents are clearly  modified [29]. The field
$\sigma_1$ is on special
footing because it always appears through its energy momentum tensor so
it
can be replaced with an effective $T_{eff}$ of any conformal field
theory
as long as it has the same value of the central charge. If one
switches-off
$all$ the matter fields as if one had effectively a "critical"
string comprised only on the Liouville sector, one could  replace :
$$T(\sigma_1) =-1/2~(\partial_z \sigma_1)^2 -\alpha_o \partial^2_z
\sigma_1. \eqno
(20)$$
with a central value
$$c=1+12(\alpha_o)^2. \eqno (21)$$
by an effective stress energy tensor associated with $D$ worldsheet
scalars, $X^\mu$, with a background charge vector $\alpha_\mu$ :
$$T_{eff}=-1/2~\partial X_\mu \partial X^\mu -\alpha_\mu.\partial^2
X^\mu.
\eqno (22)$$
so that  a $c_{eff} =D+12\alpha_\mu.\alpha^\mu $ equal to (21) will do
the job. This will be sufficient
to ensure closure of the $W_N$ algebra (after switching-off the matter
sector) once one $includes$ the extra fields
$\sigma_2,\sigma_3.......$
In  [29] it was
emphasized that noncritical strings involve $two$ copies of the $W_N$
algebra. One for the matter sector and one for the Liouville sector.
Since
$W_N$ is nonlinear one cannot naively add two realizations of it and
obtain
a third realization. Nevertheless there are is a way in which this is
possible [29]. In any case in order to get a nested sum of nilpotent
BRST operators,
{\bf $Q^n_N$}, one needs to have  $all$ the matter fields, $\phi_k$;
the
scalars ( in the nested basis)
$\sigma_{n-1},.......\sigma_{N-1},$ of the Liouville sector and the
ghost
and antighost fields of the spin-$n,n+1,......N$ symmetries where $n$
ranges between $2$ and $N$.  Central charges were computed for each set
of the nested set of stress energy tensors, $T^n_N$  depending on the
above fields which appear in the construction of the $Q^n_N$ BRST
charges.  The important point is that the value of $x$ can be chosen at
will. Later we shall see the connection that this choice has with Toda
field theory and constrained $SL(N,R)$ WZWN models at level $k$.
Therefore, the effective central charge in the noncritical $W_N$ string
case is now  $c_{m_o}+1+12(\alpha_o)^2$ in contradistinction to the
critical case
$1+12 (\alpha)^2$ where we define :
$$c_{eff} \equiv 1-12x^2 =(1-12x^2_o)+c_{m_o}.
{}~-12x^2=-12x^2_o+c_{m_o}. \eqno (23)$$
with $c_{m_o}$ given by (19). Thus in this way one has coupled
$W_N$ matter to $W_N$ gravity consistently where the Liouville modes
have
now been switched-on  and a unitary noncritical  $W_N$ string spectrum
can be constructed when the central
charge equals :
$c_{m_o}+1+12(\alpha_o)^2$.

As stated earlier the induced covariant $W_N$ action- which is
tantamount of
coupling $W_N$ matter to $W_N$ gravity- in the conformal gauge reduces
to a Toda action [32]. The constraints following after the gauge fixing
of the
$W_N$ generalizations of the Beltrami differentials form a closed $W_N$
classical algebra. The $W_N$ currents are comprised of a "matter"
sector
$and$ a Toda sector. Quantization can be achieved by constructing a
nilpotent
BRST operator in the lines of [29]. This is the connection between the
membrane solutions (8c) in terms of (10) via (9a,9b) and the solutions
to
the continuous Toda equation.

If one chooses the value  for $x^2_o$ given in (18)  with the
particular value of $q =N+1$ , one gets that
$c=1+12(\alpha_o)^2=1-12x^2_o$ becomes in the
$N\rightarrow \infty$ limit [29] : $1-12x^2_o\rightarrow
25. $ (we are setting $q=N+1$).
Since we do not wish to break the target spacetime Lorentz invariance
one
$cannot$ have background charges for the $D~X^\mu$ coordinates.
Therefore
if one chooses for  the
effective stress energy tensor the $D~X^\mu$ coordinates
without background charges
one has in this case that an effective central charge  given by (23):
$c_{m_o}+1-12x_o^2=c_{m_o}+25$ can be achieved by having :

$$D =c_{m_o}+25=1-12x^2. \eqno (24)$$
$c_{m_o}$ is explicitly given by eq-(19) which in the case that $q=N+1$
it
becomes :
$$c_{m_0} =2{N-1\over {N-2}}\rightarrow 2. \eqno (25)$$
in the $N\rightarrow \infty$ limit.
Therefore,  in the $N\rightarrow \infty$ limit, (24,25) become :
$$D=2+25 =27 \Rightarrow x^2 =-13/6 \eqno (26)$$
and the $D=27$ $X^\mu$ scalars without background charges could be seen
as the membrane's embedding coordinates.

The value for the total central charge of the matter sector is $c_m
=c_{m_o} +{1\over 24}$ after a zeta function regularization. The
central charge for the Liouville sector is $c_L =-4-{1\over 24}$. The
value of $c_m$ (after a regularisation) corresponds to the central
charge of the first unitary minimal model of $WA_{n-1}$ after $n$ is
analytically continued to a negative value of $n=-146 \Rightarrow
2(n-1)/(n+2) =2+1/24 $. The value of $c_L$ does not correspond to a
minimal model but nevertheless corresponds to a very special value of
$c$ where the $WA_{n-1}$ algebra truncates to that of the $W$ algebra
associated with noncompact coset models [40] :
$$WA_{n-1} \Rightarrow W(2,3,4,5)\sim {sl(2,R)_n \over {U(1)}}. \eqno
(26b)$$
This occurs at the value $c(n)=2(1-2n)/(n-2)=-4-1/24$ for $n=146$.  Two
other values for $c(n)$ are possible , in particular the first unitary
minimal model.  One should not confuse $c_{eff} $ with $c_m$ and $x^2$
with $x^2_o$.

If  the BRST quantization of
the continuous Toda action is devoid of $W_{\infty}$ anomalies the net
central charge of the matter plus Toda sector must equal to $-2$ which
is
the regularized value obtained by [34]. Saveliev
[37] gave a highly nontrivial realization of the $W_{\infty}$ algebra
in terms of the $3D$ continous Toda field $\Phi (z,{\bar z},t)$
as we discussed in the previous section. One has to $check$ that indeed
such a realization of the $W_{\infty}$ algebra, after a BRST
quantization
or another quantization scheme, does have the correct  central charge
in
order to have $c_m+c_L$ equal to $-2$. We recall once more the fact
that
{32] have shown that the Toda action is the action one gets after one
couples $W_N$ matter to $W_N$ gravity and the conformal gauge is
chosen.
It is well known to the experts by now that [41] Quantum Toda theories
are conformally invariant and the conformally improved stress energy
tensor obeys a Virasoro algebra with an $adjustable$ central charge
which depends on the value of the coupling constant $\beta$ appearing
in the exponential potential.  This value for the central charge
$c(\beta)$ coincides precisely with the one obtain from a Quantum
Drinfeld-Sokolov reduction of the $SL(N,R)$ Kac-Moody algebra at the
level $k$.  The value of the coupling is :
$$\beta ={1\over \sqrt {k+N}}.~c(\beta) =(N-1) -12|\beta \rho -1/\beta
\rho ^v|^2. \eqno (26c)$$
where $\rho,\rho^v$ are the Weyl vectors of the (dual) $A_N$ Lie
algebra. [41] . We see that indeed (26c) has the same form as (17)
where the value of $x^2$ is related to the coupling $\beta$ :
$$|x^2| =13/6\sim (1/\sqrt {k+N} -\sqrt
{k+N})^2.~k=-\infty.~k+N=constant. \eqno (26d)$$
So we learn that in the $N\rightarrow \infty$ limit one must have
$k=-\infty$ such as  $k+N=constant$ in order to obey (26d). The check
now is to evaluate the central charge directly from Saveliev's
realization of the $W_{\infty}$ algebra in terms of the $3D$ continuous
Toda theory. From eqs-(9a,9b) one has that the continuous Toda field
$\Phi$ is a function of the YM potentials, $A_\mu$, which play the role
of coordinates :$A_\mu \rightarrow X_\mu$ so that $\Phi =\Phi[X_\mu]$.
Upon computing the operator products of the $W_{\infty}$ currents in
terms of $\Phi$ one can reexpress them in terms of OPE involving the
$X_\mu$ coordinates which from the point of view of a three-dimensional
observer are just $3D$ scalar fields.  Therefore, in order to be
$selfconsistent$  the central charge obtained from such a procedure
must $match$ the central charge obtained from the value of $c(\beta)$
above in eqs-(26). This selfconsistency in the matching of these
central charges would be an indication that $x^2$ is in fact equal to
$-13/6$ and that the zeta function regularisation was indeed consistent
with the Saveliev's realization. The value of $x^2=-13/6$ is the
required value so that $c_{eff} =27$. Fixing $x^2$ yields the coupling
constant $\beta$ in (26c, 26d).
Notice that:
$$D-2=25=26-[1-{6\over {q(q+1)}}].\eqno (27)$$
in the $N\rightarrow \infty$ limit. So a connection with the unitary
Virasoro
minimal models is also established for those values of $q=N+1$.

In the supersymmetric case one could again establish the connection
with the
unitary Virasoro superconformal minimal models. In the $N\rightarrow
\infty$ limit
the value of the central charge becomes $c(superconformal)\rightarrow
3/2$.
Since $10$ is the citical dimension of the superstring the value of the
central charge when one has ten worldshhet scalars and ten fermions
is $10+10/2 =30/2$. Therefore in order to have an effective "critical
"super
$W_{\infty}$ string one has for the supersymmetric analog of the right
hand
side of (27) the value of :
$$10(1+1/2)-c_{superconformal}=30/2-3/2=27/2. \eqno (28)$$

The super Liouville sector is comprised of the infinite number of super
Toda
fields present in the continous super $SL(\infty)$ Toda action
and must have a total central charge $c_L$ so that $c_m+c_L$ is equal
to $-3$ which is the required
value to have an anomaly-free "matter" realization of the quantum super
$W_{\infty}$
algebra [34]. Going through the same reasoning as in the bosonic case
[29]
and choosing the appropriate value for the arbitrary parameter $x^2$
in
order to make contact with the bosonic sector of the  unitary
minimal models of the super $W_N$ algebra ones gets for the central
charge
of the matter sector of the super $W_N$ algebra the value of $c_{m_o}=3
=2(1+1/2)$.

Having $D~X^\mu$ and  $D~\psi^\mu$
(anticommuting spacetime vectors and worldsheet spinors) without
background
charges yields a central charge equal to $D+D/2=3D/2$.
Equating now :
$$3D/2 =c_{m_o}+[30/2-3/2]=3+27/2 \Rightarrow D=11. \eqno (29)$$
This is the expected critical dimension for the supermembrane. So one
obtains the expected critical dimensions for the (super) membrane if
one
adjoins a $q=N+1$
unitary minimal model of the $W_N$
algebra in the $N\rightarrow \infty$ limit to a critical $W_{\infty}$
string spectrum. The same goes for the super $W_N$ algebra.

The issue of global anomalies remains to be settled. Global anomalies
in the $E_
8\times
E_8$ and $SO(32)$ superstring theories in arbitrary target spacetimes
were discussed in [20] . Cancellation of global anomalies in the
$E_8$ theory results in
constraints on the topology of $M$ . For the $SO(32)$ it seems unlikely
that
any nontrivial manifold leads to an anomaly free theory.
Since the exceptional algebra $E_8$ is intimately tied up with the
octonion
division algebra which is so essential to formulate a SDYM in $R^8$
[1,9]
refered above, prior to the reduction to $R^4$, it seems that an
anomaly-free
supermembrane can only propagate in those special manifolds (after one
spatial
dimensional reduction from $D=11 \rightarrow D=10$) discussed in [20].

As far as Lorentz
anomalies is concerned we have seen how the cancellation of
$W_{\infty}$ anomalies
for the noncritical $W_{\infty}$ strings ( the {\bf net} value of $c=0$
)
via a BRST analysis might yield
information of the supermembrane's critical
dimension. Conformally-invariant Spinning membrane actions
have been given in [21] in a nonpolynomial form,  and in polynomial
form [22].
. The polynomial form [22] was supersymmetric solely under the $Q$
super
symmetry of the superconformal algebra in $D=3$ and was $not$ invariant
under $S$ supersymmetry nor $K$ symmetry ( conformal boosts). It was
invariant
under translations, Lorentz and dilations and the subalgebra comprised
of
these invariance-generators does in fact $close$.
The action [22], in effect, was a $3D$-conformally invariant analog of
the spinning
string with the main difference that one needs the explicit presence of
the
$3D$-gauge field of dilations, $b_\mu$ which depends on the matter
fields (the membrane's

coordinates) in a highly complicated manner due to the
presence of the $quartic$ derivative terms. It is unfortunate that we
have encou
ntered
so many difficulties in having many readers accept this point.

It is essential to study the plausible existence of conformal
anomalies
and see if constraints on the dimension of the target spacetime can be
obtained as they do for the ordinary string.
These latter spinning membrane actions are the supersymmetric extension
of the generalization of the
$SU(2)$ chiral models in three spacetime
dimensions discussed in  [23]. Finite, stable energy and static
kink solutions carrying a topologically
conserved charge were found without the usual scaling instability
problems
present in $D=2$ (Derrick's theorem). This required the choice of the
nonpolynomial
action [21] whereas the polynomial action [22] bears  a
strong resemblance with Skyrme's model.
A simple count of the transverse dimensions of the supermembrane in
$4,5,7,11$
dimensions suggests that because $11-3 =8 =(3)^2 -1 \Rightarrow $ one
can incorporate these membrane's degrees of freedom into a $SU(3)$
generalization of
Skyrme's model.
The other values of $D$ do not fit.

Finally we wish to mention the importance of these  (instanton)
solutions
of the $D=11$ supermembrane  in connection to the the geometry of SYM.
Firstly,  a complexification yields the $SU^*(\infty)$
group which is isomorphic to $SL(\infty ,H)$ and hence the role of
quaternions in these theories.
All this seems to point at the fact that it is
quaternionic field theory in $D=4$ [24]
the one which  may contain many of the hidden symmetries of string
theory
and in particular
might give us the clues how to build a background-independent
string-field
theory in terms of the spaces of
all quaternionic field theories. Secondly, the abundant research into
twistor-like formulations
of null super-$p$-branes [25]  by extending the configuration space
through the addition of auxiliary spinor coordinates of the
Newman-Penrose dyades for $D=4$, will give us important clues as to how
implement
quaternionic analyticity into the picture. ( in terms of quaternionic
twistors).

And, finally, the notion of integrability
on light-like lines [26] is tightly connected with supertwistors
( a parameter
space of light-like lines in complex superspace ). It was such
integrability
in $D=10~N=1$ and $D=4$ for $N$ greater than $ 2 $ super YM theory
that led to the on-shell equations of motion.
Per example : $N=1~D=6$
SYM in $complexified$-superspace can be written down in terms of
$SL(2,H)\sim SU^*(4)$ spinors.
A dim reduction to $D=4$ implies a fourth-folding in terms of the
number of $real$
supersymmetries and one has then  a $D=4~N=4~SYM$ which many believe it
is finite theory
at all loops [27]. Since in $D=10~SYM$ the number of transverse degrees
of freedom
is eight which is the same   as those of the supermembrane it would be
interesting
if extending the notion of light-like integrability in loop superspaces
or
their generalization to higher-dimensional loops,
yields exact on-shell solutions to the supermembrane's equations of
motion.
i.e; Instead of light-like lines one has null-$p$-branes directions in
higher
dimensional loop spaces.
After all it has been known for some time that SDYM amounts to an
integrability condition. The construction of higher dimensional loop
algebras with central extensions and their relevance to extended
objects
has been given by Goteborg group [42].
We hope that the super-Toda molecule equation is an
alternative to the SGQMM. This is the main conclusion of this work.

\section{Acknowledgements}

We thank M.V.Saveliev for discussions, references and suggestions in
connection
to the
quantization of the continuous Toda equation. To Jurgen Rapp for his
help and ho
spitality at the KFA, Julich, Germany
where this work was completed.;and to Bob Murray in Austin,Texas. Also,
to Tanya Stark for her support and warmth in Koln, Germany.

\section{Appendix}
We shall give in this appendix the continuum limit of (10). For
simplicity we shall only work with the bosonic sector. It is given by
$${-\partial^2 \Psi \over {\partial t^2}} =\int~K(y,y')exp[\Psi
(t,y')]dy'.
\eqno (A-1)$$
The continuum field $\Psi (t,y=y_0 +l\epsilon ) \equiv \epsilon \Phi_l
(t)$.
Where one partitions the interval $[y_0,y_f]$ in $N$ intervals and
performs the
linear interpolation such as :
$${\Psi (t,y_f) -\Psi (t,y_0) \over {\Phi_N -\Phi_1}} \sim {y_f -y_0
\over N} =\epsilon. \eqno (A-2)$$

In this fashion, when $\epsilon \rightarrow 0$ the continuum field
$\Psi (t,y_0 ) =\Psi (t,y_f) \rightarrow 0$ so that $\Phi_1
;\Phi_\infty$ are
well defined. The continuum Cartan matrix is :
$K(y,y') =\epsilon K_{ll'}$
where $y_0+l\epsilon \le y \le y_0 +(l+1)\epsilon$ and
$y_0+l'\epsilon \le y' \le y_0 +(l'+1)\epsilon$. Another way of
recasting the
continuum Cartan matrix is :$ \delta '' (y-y')$ (we thank I. Bakas for
pointing

this out to us) so that (A-1) becomes :
$$-\partial^2_{t^2} \Psi =\partial^2_{y^2}~exp \Psi.\eqno (A-3)$$

In solving equation (A-3) one could recurr to
a dimensional reduction from $3\rightarrow 2$ of the $sl(\infty)$ Toda
equation :
given by Leznov and Saveliev :
$$ \partial_z \partial_{\bar z} u =-\partial^2_t (e^u). \eqno (A-4)$$
where $u(z,{\bar z},t)$ and $z=x+iy.~{\bar z} =x-iy$.
Therefore, the solutions to the Toda molecule equation are tied-up to
Killing
symmetry reductions of $D=4~SDG$.

\section{References}
1. T. A. Ivanova, A.D. Popov : Jour. Math. Phys. {\bf 34} no.2 (1993)
674.

2.R.S. Ward : Phys. Lett. {\bf A 112} (1985) 3.

S. Rouhani : Phys. Lett. {\bf A 104} (1984) 7.

3.M. Duff : Class. Quantum Grav. {\bf 6} (1989) 1577 and references
therein.

4.C. Castro : Phys. Lett. {\bf B 288} (1992) 291.

5.M.P. Grabowski , C.H. Tze : "On the Octonionic Nahm equations and
Self Dual
Membranes in 9 Dimensions " VPI-IHEP-92-9 preprint .

6.A.D. Popov : Modern Phys. Lett {\bf A 7} no.23 (1992) 2077

and references therein. JETP Lett. {\bf 55} (1992) 259.

A. N. Leznov and M. V. Saveliev  : Comm. Math. Phys {\bf 74} (1980)
111.

N. Ganoulis, P. Goddard and D. Olive : Nucl. Phys. {\bf B 205} (1982)
601.

7.E.G. Floratos , G.K. Leontaris : Phys. Lett. {\bf B 223} (1989) 153.

8. C.R. Gilson, I. Martin , A. Restuccia, J.G.Taylor : Comm. Math.
Phys. {\bf
107} (1986) 377.

9.J. Harvey, A. Strominger : Phys. Rev. Lett. {\bf 66} no.5 (1991) 549.

10.M.J. Duff, J.X. Lu : Class.Quantum Grav. {\bf 9} (1992) 1.

11.T. Kugo, P. Townsend : Nuc. Phys. {\bf B 221} (1983) 357.

12.B. DeWit, M. Lusscher and H. Nicolai : Nuc. Phys. {\bf B 320} (1989)
135.

13.B. Simon : Ann. Phys. {\bf 146} (1983) 209.

14.R. S. Ward : "Twistors in Mathematics and Physics ". LNSLMS vol. 156
(1990)

Cambridge Univ. Press. T. Bailey and R. Baston, editors.

15. C. Castro : Jour. Math. Phys. {\bf 35} no.6 (1994) 3013.

C. Castro : " Nonlinear $W_{\infty}$ Algebras from Nonlinear Integrable
Deformations of Self Dual Gravity " I.A.E.C-4-94 preprint.

16. Q.H. Park : Int. Journ. Mod. Phys. {\bf A 6} (1991) 1415.

I. Bakas :Comm. Math. Phys. {\bf 134} (1990) 487.

17. I. Bakas, B. Khesin, E. Kiritsis : Comm. Math. Phys. {\bf 151}
(1993) 233.

D.B. Fairlie, J. Nuyts : Comm. Math. Phys. {\bf 134} (1990) 413.
18. K. Takasaki, T. Takebe  :

Letters of Math. Phys. {\bf 23} (1991) 205.

19. S. Wu and Y. Yu :Jour. Math. Phys. {\bf 34} (1993) 5851. {\bf ibid}
5872.

20. T. P. Killingback : Class. Quantum Grav {\bf 5} (1988) 1169.

21. U.Lindstrom and M. Rocek : Phys. Lett. {\bf B 218 } (1988) 207.

22. C. Castro : To appear in the International Journal of Groups in
Phys.
{\bf vol 2 } (1994).

23. M. Duff, S. Deser, C. Isham : Nucl. Phys. {\bf B 114} (1976) 29.

24. F. Gursey and H.C. Tze : Ann. Phys {\bf 128} (1980) 29.

and in Lett. Math. Phys. {\bf 8} (1984) 387.

25.I. A. Bandos and A.A. Zheltukhin : Int. Jour. Mod. Phys. {\bf A 8}
no. 6

(1993) 1081.

26. E. Witten : Phys. Lett. {\bf B 77} (1978) 394.

J. Harnard, S. Schnider : Comm. Math. Phys. {\bf 106} 183.

27. M.Grisaru, W. Siegel : Nucl. Phys. {\bf B 201} (1982) 292.

28. C. N. Pope : " A Review of $W_N$ strings" CTP-TAMU-30-92 preprint.

K. Yamishi : Phys. Lett. {\bf B 266} (1991) 370.

29. E. Bergshoeff, H.J. Boonstra, S.Panda and M. de Roo : Nucl. Phys.
{\bf B 411}(1994) 717.
E.Bergshoeff, J.de Boer, M.Roo and T.Tjin : Nucl. Phys. {\bf B
420}(1994) 379.

30. J.L Gervais, J. Schnittger : " The Many Faces of Quantum Liouville
Exponentials "

LPTENS-93-30 preprint.

31.L. Frappat, E. Ragoucy and P. Sorba : " $W$ (Super) Algebras from
constrained WZW models:

A Group Theoretical Classification". ENSLAPP-AL-391/92 preprint.

32. J. de Boer, J. Goeree : Phys. Lett.{\bf B 274} (1992) 289.

 Nucl. Phys {\bf 405 B}(1993) 669.

33.X. Shen :" $W_\infty$ and String Theory ". CERN-TH-6404/92 preprint.

34. C.N.Pope,L.J.Romans and X. Shen : Phys- Lett. {\bf B 254} (1991)
401.

35. C.N.Pope : "A Review of $W_N$ Strings " CTP-TAMU-30/92 preprint.

36. S.R.Das, A. Dhar and S.K.Rama : Int. Jour. Mod. Phys. {\bf A7}no.10
(1992) 2293.

37.M.V.Saveliev : Theor. Math. Phys. {\bf vol. 92} no.3 (1992) 457.

38.A.N.Leznov, M.V.Saveliev : "Group Theoretical Methods of Integration
of
Nonlinear Dynamical systems" Birkhauser, 1992.

A.N. Leznov and M.V. Saveliev Acta. Appl. Math {\bf 16} (1989)1-74.

39.S.Odake : Int. Jour. Mod. Phys {\bf A7} (1992) 6339.
V.Kac and O.Radul : Comm. Phys. {\bf 157} (1993) 429.

40.R. Blumenhagen, W. Eholzer, A. Honecker, K. Hornfeck and R. Hubel :
Phys. Lett. {\bf B 332} (1994) 51-60.

41. P. Bouwknegt and K. Schouetens : Phys. Reports 223 (1993) 183.

42.M. Cederwall, G. Ferretti, B. Nilsson and A. Westerberg :
Goteborg-ITP-93-37 and hepth-9401027.

\enddocument